\documentclass[tightenlines,superscriptaddress,eqsecnum,floats,nofootinbib,showpacs]
{revtex4}
\usepackage{amssymb}
\usepackage{stmaryrd}
\usepackage{amsmath}
\usepackage{amsfonts}
\usepackage{mathrsfs}
\usepackage{hyperref}
\usepackage{amsmath,amssymb,amsfonts}
\usepackage{graphicx}

\def\be{\begin{equation}}
\def\ee{\end{equation}}
\def\ba{\begin{eqnarray}}
\def\ea{\end{eqnarray}}
\def\nn{\nonumber}

\makeatletter
    
    \newcommand{\Rmnum}[1]{\expandafter\@slowromancap\romannumeral #1@}
    \makeatother

\newcommand{\kt}{{\tilde{K'}}}
\newcommand{\R}{\mathcal {R}}

\newcommand{\ints}{{\int_\Sigma}}

\begin{document}

\title{Connection dynamics of higher dimensional scalar-tensor theories of gravity}

\author{Yu Han}\email{hanyu@mail.bnu.edu.cn}
\author{Yongge Ma\footnote{Corresponding author}}\email{mayg@bnu.edu.cn}

\affiliation{Department of Physics, Beijing Normal University,Beijing 100875,China}

\author{Xiangdong Zhang}\email{zhangxiangdong@mail.bnu.edu.cn}

\affiliation{Department of Physics, South China University of Technology,Guangzhou 510641,China}

\begin{abstract}
The scalar-tensor theories of gravity in spacetime dimensions $(D+1)>2$ are studied. By performing Hamiltonian analysis, we obtain the geometrical dynamics of the theories from their Lagrangian. The Hamiltonian formalism indicates that the theories are naturally divided into two sectors by the coupling parameter $\omega$. The Hamiltonian structures in both sectors are similar to the corresponding structures of 4-dimensional cases. It turns out that, similar to the case of general relativity, there is also a symplectic reduction from the canonical structure of $so(D+1)$ Yang-Mills theories coupled to the scalar field to the canonical structure of the geometrical scalar-tensor theories. Therefore the non-perturbative loop quantum gravity techniques can also be applied to the scalar-tensor theories in $D+1$ dimensions based on their connection-dynamical formalism.

\end{abstract}


\maketitle

\section{Introduction}
Ever since 1998, a few independent astronomic observations
strongly suggested that our Universe is currently undergoing a period of acceleration \cite{Fr08}. This causes
the ``dark energy" problem in the framework of general relativity (GR). While a positive cosmological
constant $\Lambda$ could be employed to explain the acceleration, the observed value of $\Lambda$ is unexpectedly much smaller than any theoretical estimation. Therefore it is reasonable to consider the possibility that GR is not a valid theory of gravity on galactic or cosmological scale. For this reason, as well as some non-trivial tests on gravity beyond GR \cite{Will, Will05, Will051},  modified gravity theories have received increased attention recently. Among various alternative models, the typical candidate is the so-called $f(\R)$ theory \cite{So}. Besides $f(\R)$ theories, Brans-Dicke theory of gravity which was first proposed by Brans and Dicke in 1961 and compatible with Mach's principle \cite{BD} also caught much attention. In this theory, a scalar field representing the varying ``gravitational constant" is non-minimally coupled to the scalar curvature. To interpret the observational results within the framework of a broad class of theories, the Brans-Dicke theory was generalized by Bergmann \cite{bergmann} and Wagoner \cite{wagoner} to scalar-tensor theories (STT). The scalar field in STT of gravity is expected to account for the mysterious ``dark energy", since it can naturally lead to cosmological acceleration in certain models (see e.g. \cite{Boiss00, Banerjee, Boiss11, Qiang}). In particular, the current acceleration of the Universe can be naturally obtained in 5-dimensional Brans-Dicke theory without fine-tuning of the coupling parameter \cite{Qiang, Qiang2}. Moreover, some models of STT of gravity may also account for the ``dark matter" problem \cite{lee, catena, kim}, which was revealed by the observed rotation curve of galaxy clusters. Besides, scalar-tensor modifications of GR have also become very popular as the low-energy and effective limit in unification schemes such as bosonic string theory (see e. g. \cite{tayler, maeda, damour}). It should be noted that the general scalar-tensor theory can include both metric $f(\R)$ theories and Palatini $f(\R)$ theories as special sectors with different coupling parameter $\omega$, while the original Brans-Dicke theory is the particular case of constant $\omega$ and vanishing potential.

On the other hand, during the past several decades, seeking for a quantum theory of gravity has been a rather active field. Among various kinds of programmes, loop quantum gravity (LQG), a background independent approach to quantize GR, has been widely investigated \cite{Ro04, Th07, As04, Ma07}. Surprisingly, as a non-renormalizable theory from the view of perturbative quantum field theory, GR can be non-perturbatively quantized by the loop quantization procedure. The loop quantization programme heavily relies on the connection-dynamical formulation of GR, which requires a Poisson self-commuting connection variable and a corresponding compact gauge group. While the approach to formulate the connection dynamics was originally restricted to 4-dimensional GR, it can also be generalized to 4-dimensional $f(R)$ theories and general STT \cite{Zh11, Zh11b, Zh11c,Zh13b,ZAM,Zhou13}. It is shown in a cosmological model that the quantization of STT in Einstein and Jordan frames are not equivalent to each other \cite{AMZ}.
However, modern theoretical research indicates that we might live in a universe with spacetime dimension $D+1>4$. Thus one is naturally led to ask whether higher dimensional gravity theories can be formulated as gauge theories with connection dynamics. Recently, in a series of seminal articles \cite{Th03,Th04}, Bodendorfer, Thiemann and Thurn successfully developed an approach to formulate the connection dynamics for GR as well as supergravity theories in higher dimensions \cite{Nb11, At11,At11a}. Taking account of the cosmological and astrophysical significance, it is desirable to study if the connection-dynamical formalism also exists for STT in arbitrary dimensions. In this paper we will give an affirmative answer to this question. Our results can serve as the starting-point for the non-perturbative loop quantization of STT in higher dimensions.

This paper is organized as follows. In section 2, we formulate the Hamiltonian analysis of $(D+1)$-dimensional $(D>1)$ STT in terms of ADM variables. In section 3, we first give a brief review of the new variables and connection dynamics of GR in $D+1$ dimensions. Then we show how to obtain the ADM variables from the new connection variables of $(D+1)$-dimensional STT by symplectic reduction. We will write out the explicit form of the four different constraints and prove that they indeed form a first-class constraint system when $\omega(\phi)\neq-\frac{D}{D-1}$. For the special case when $\omega(\phi)=-\frac{D}{D-1}$, a new constraint generating spacetime conformal transformations is found. The five different constraints also form a first-class system. We summarize our results in the last section. The detailed calculations of several Poisson brackets will be given in appendix A . Throughout the paper, we use Greek alphabet $\mu,\nu,...$ for
spacetime indices, Latin alphabet $a, b, c, . . . , $ for spatial indices,
and $\textit{I}, \textit{J}, \textit{K}, . . . ,$ for internal indices.

\section{Hamiltonian analysis}
In the vacuum case, the general action of $(D+1)$-dimensional scalar-tensor theories reads:
\ba
S[g,\phi]&=&\int_\Sigma
d^{D+1}x\sqrt{-g}\Big[\frac12\Big(\phi\R-\frac{\omega(\phi)}{\phi}(\partial_\mu\phi)\partial^\mu\phi\Big)-V(\phi)\Big],\label{action}\ea
where we set $8\pi G=1$, the coupling parameter $\omega(\phi)$ and potential $V(\phi)$ can be arbitrary functions of the scalar field.  The field equations read
\ba G_{\mu\nu}&=&\frac{1}{\phi}(\nabla_\mu\nabla_\nu\phi-g_{\mu\nu}\nabla^\sigma\nabla_\sigma\phi)
+\frac{\omega(\phi)}{\phi^2}[(\nabla_\mu\phi)\nabla_\nu\phi-\frac12g_{\mu\nu}(\nabla\phi)^2]-g_{\mu\nu}\frac{V(\phi)}{\phi}, \label{01}
\\
\phi\R&=&-2\omega(\phi)\nabla^\sigma\nabla_\sigma\phi+(\frac{\omega(\phi)}{\phi}-\omega'(\phi))(\nabla_\mu\phi)\nabla^\mu\phi+2\phi V'(\phi), \label{02}\ea
where the prime denotes the derivative  with respect to $\phi$. By performing $D+1$ decomposition, the Lagrangian density in Eq.(\ref{action}) becomes
\ba
\mathcal{L}=&&\frac12N\sqrt{h}\phi(R{}^{(D)}+K_{ab}K^{ab}-K^2-2\frac{V}{\phi})-\sqrt{h}K[\dot{\phi}-(N^a\partial_a\phi)]+\frac{\sqrt{h}\omega(\phi)}{2N\phi}[(\dot{\phi}-N^a\partial_a\phi)^2-N^2h^{ab}(\partial_a \phi) \partial_b\phi]\nn\\
&&+\sqrt{h}h^{ab}(\partial_a\phi) \partial_b N,
\ea
where $K_{ab}$ denotes the
extrinsic curvature of the $D$-dimensional spatial hypersurface $\Sigma$, $R^{(D)}$ is the Ricci scalar of the $D$-metric
$h_{ab}$, $N^a$ and $N$ are respectively the shift vector and lapse function. The configuration variables in this theory are $(h_{ab},\phi)$,  their
conjugate momenta are defined by

\ba
 \Pi^{ab}&=&\frac{\partial\mathcal
{L}}{\partial\dot{h}_{ab}}=-\sqrt{h}[\frac{\phi}{2}(K^{ab}-Kh^{ab})+\frac{h^{ab}}{2N}(\dot{\phi}-N^c\partial_c\phi)],  \label{pab1}\\
\pi&=&\frac{\partial\mathcal
{L}}{\partial\dot{\phi}}=-\sqrt{h}[K-\frac{\omega(\phi)}{N\phi}(\dot{\phi}-N^c\partial_c\phi)]. \label{pi}
\ea
The combination of Eq.(\ref{pab1}) and Eq.(\ref{pi}) yields
\ba
K_{ab}
&=&\frac{2(\Pi_{ab}-\frac{1+\omega}{D+(D-1)\omega}\Pi h_{ab})}{\phi\sqrt{h}}-\frac{\pi
h_{ab}}{(D+(D-1)\omega)\sqrt{h}},\label{excur}
\ea
and
\ba
 \Pi-\frac{D-1}{2}\phi\pi=\frac{(D-1)\sqrt{h}}{2N}(N^a\partial_a\phi-\dot{\phi})[\omega(\phi)+\frac{D}{D-1}], \label{Sconstraint}
\ea
where $\Pi\equiv h_{ab}\Pi^{ab}$. It is easy to see
that $C\equiv \Pi-\frac{D-1}{2}\phi\pi$ is constrained to vanish when $\omega(\phi)=-\frac{D}{D-1}$, which corresponds to the case of  Palatini $f(R)$ gravity
in $D+1$ dimensions. So this theory is naturally marked off into two different sectors by
$\omega(\phi)\neq-\frac{D}{D-1}$ and $\omega(\phi)=-\frac{D}{D-1}$. In the following we perform the constraint analysis of these two sectors separately.

\subsection{Sector of $\omega(\phi)\neq-\frac{D}{D-1}$ }
In the case $\omega(\phi)\neq -\frac{D}{D-1}$,  the total Hamiltonian of STT can
be expressed as a liner combination of smeared constraints as
\ba H_{tot}=H[\overrightarrow{N}]+H[N], \label{htotal} \ea
where
the smeared diffeomorphism and Hamiltonian constraints read
as follows
\ba H[\overrightarrow{N}]&=&\int_\Sigma d^DxN^aH_a =\int_\Sigma
d^DxN^a\left(-2D^b(\Pi_{ab})+\pi\partial_a\phi\right), \label{dc}\\
H[N]&=&\int_\Sigma d^DxNH \nn\\
&=&\int_\Sigma
d^DxN\Big[\frac2{\sqrt{h}}\Big(\frac{\Pi_{ab}\Pi^{ab}-\frac{1}{D-1}\Pi^2}{\phi}+\frac{(\Pi-\frac{D-1}{2}\phi\pi)^2}{\phi(D-1)(D+(D-1)\omega)}\Big)\nn\\
&&\quad\qquad+\frac12\sqrt{h}\Big(-\phi R{}^{(D)}+\frac{\omega(\phi)}{\phi}(D_a\phi)
D^a\phi+2D_aD^a\phi+2 V(\phi)\Big)\Big],\label{hc}\ea
where $D_a$ denote the spatial covariant derivative compatible with $h_{ab}$. By the standard Poisson brackets
\ba
\{h_{ab}(x), \Pi^{cd}(y)\}&=&\delta^{(c}_a\delta^{d)}_b\delta^D(x, y), \nn\\
\{\phi(x), \pi(y)\}&=&\delta^D(x, y),  \label{poission}\ea
lengthy calculations show that the constraints (\ref{dc})
and (\ref{hc}) comprise a first-class system similar to GR as:
\ba \{H[\overrightarrow{N}], H[\overrightarrow{N}^\prime]\}&=&H([\overrightarrow{N}, \overrightarrow{N}^\prime]),  \nn\\
\{H[M], H[\overrightarrow{N}]\}&=&H[\mathcal
{-L}_{\overrightarrow{N}}M],  \nn\\
\{H[N], H[M]\}&=&H_a[ND^aM-MD^aN]. \ea

Next we show that the evolution equations of the canonical variables is
consistent with the field equations (\ref{01}) and (\ref{02}). The evolution equations
can be derived by calculating their Poisson
brackets with the total Hamiltonian (\ref{htotal}). Firstly, it is obvious that the evolution equation of $h_{ab}$ is just the
definition of $K_{ab}$. Secondly, the evolution equation of $\Pi_{ab}$ reads
\ba
\dot{\Pi}_{ab}&=&\frac N4\phi \sqrt{h}\Big(h_{ab} R{}^{(D)}-2 R^{(D)}_{ab}\Big)+\frac{Nh_{ab}}{\sqrt{h}}\Big(\frac{\Pi_{cd}\Pi^{cd}-\frac{1}{D-1}\Pi^2}{\phi}+\frac{(\Pi-\frac{D-1}{2}\phi\pi)^2}{(D+1)\phi(D+(D-1)\omega)}\Big)
+\frac{4N}{\sqrt{h}}\Big(\frac{\Pi_{ac}\Pi_b{}^c-\frac{1}{D-1}\Pi\Pi_{ab}}{\phi}\nn\\&&+\frac{(\Pi-\frac{D-1}{2}\phi\pi)\Pi_{ab}}{(D-1)\phi(D+(D-1)\omega)}\Big)
-\frac
{N\omega}{4\phi}\sqrt{h}\Big(h_{ab}(D_c\phi)D^c\phi-2(D_a\phi)D_b\phi\Big)
-\frac{N\sqrt{h}}{2}\Big(2h_{ab}D_cD^c\phi+D_{(a}D_{b)}\phi \Big)
\nn\\
&&+\frac{\sqrt{h}}{2}\phi (D_{(a}D_{b)}N-h_{ab}D_cD^cN)-\sqrt{h}h_{ab}(D_c\phi) D^c N+2\Pi_{c(a}D^cN_{b)}+D_c(\Pi_{ab}N^c)
-\frac12N\sqrt{h}h_{ab} V(\phi). \label{pab} \ea
Using Eq. (\ref{excur}), we can derive the evolution equation of the extrinsic
curvature as:
\ba \dot{K}_{ab}&=&-N(R^{(D)}_{ab}-2K_{ac}K^c_b+KK_{ab})+\frac{N}{\phi}\Big(D_aD_b\phi+\frac{\omega}{\phi}(D_a\phi)
D_b\phi\Big)\nn\\
&&-(\frac{\dot{\phi}}{N}-\frac{N^c \partial_c \phi}{N})K_{ab}+\mathcal
{L}_{\overrightarrow{N}}K_{ab}+\frac{Nh_{ab}}{\phi}\Big(\frac{1}{D-1}\Box\phi+\frac{2}{D-1} V(\phi)\Big). \label{dotkab}
\ea
By substituting
\ba
\mathcal{R}^{(D+1)}=R{}^{(D)}+K_{ab}K^{ab}-K^2+\frac{2}{N\sqrt{h}}[\partial_t(\sqrt{h}
K)-\partial_a(\sqrt{h}N^a
K)-\partial_a
(\sqrt{h}h^{ab}\partial_bN)],\label{D+1}
\ea
into Eq.(\ref{dotkab}),
it is straightforward to check that Eq. (\ref{dotkab}) is in accordance with Eq. (\ref{01}). Moreover, we have
\ba
\dot{\phi}=\{\phi, H_{tot}\}=\frac{2N}{(D+(D-1)\omega)\sqrt{h}}(\frac{D-1}{2}\phi\pi-\Pi)+N^a\partial_a\phi,
\ea
which is just Eq. (\ref{Sconstraint}). Finally, the time derivative of $\pi$ reads
\ba
\dot{\pi}
&=&\frac{N\sqrt{h}}{2}(R^{(D)}+K_{ab}K^{ab}-K^2)+\partial_a(N^a\pi)
-\partial_a(\sqrt{h}h^{ab}\partial_bN)-\frac{\omega\sqrt{h}}{2\phi^2N}(\dot{\phi}-N^c\partial_c\phi)^2\nn\\
&&-(\frac{N\omega\sqrt{h}}{2\phi^2}-\frac{\omega'(\phi)N\sqrt{h}}{2\phi})(D_a\phi) D^a\phi+\frac{\sqrt{h}\omega}{\phi}
(D_a N)D^a\phi
-N\sqrt{h}V'(\phi), \label{pidot} \ea
which can be proved to be equivalent to Eq. ({\ref{02}}) by using again Eq. (\ref{D+1}).
 Now we conclude that the Hamiltonian formalism of STT consists with their Lagrangian formalism when $\omega(\phi)\neq -\frac{D}{D-1}$.

\subsection{Sector of $\omega(\phi)=-\frac{D}{D-1}$ }

In this special case, Eq. (\ref{Sconstraint})
implied an extra ``conformal constraint" $C=0$. Hence the total
Hamiltonian is now expressed as
\ba H_{tot}=H[\overrightarrow{N}]+H[N]+C[\lambda], \label{htotal1}
\ea
where the definition for the smeared diffeomorphism constraint $H[\overrightarrow{N}]$ is the
same as Eq.(\ref{dc}), while the smeared Hamiltonian and conformal constraints
read respectively:
\ba
H[N]&=&\int_\Sigma
d^DxN\left[\frac2{\sqrt{h}}\Big(\frac{\Pi_{ab}\Pi^{ab}-\frac{1}{D-1}\Pi^2}{\phi}\Big)
+\frac12\sqrt{h}\Big(-\phi R{}^{(D)}-\frac{D}{(D-1)\phi}(D_a\phi)
D^a\phi+2D_aD^a\phi+2 V(\phi)\Big)\right],\nn\\\label{hc1}\\
C[\lambda]&=&\int_\Sigma d^Dx\lambda C=\int_\Sigma
d^Dx\lambda(\Pi-\frac{D-1}{2}\phi\pi). \label{sc}
\ea
Straightforward calculations give the Poisson brackets between them as:
\ba \{H[M], H[\overrightarrow{N}]\}&=&H[-\mathcal
{L}_{\overrightarrow{N}}M], \quad \{C[\lambda], H[\overrightarrow{N}]\}=C[-\mathcal
{L}_{\overrightarrow{N}}\lambda], \label{VHS}\\
\{H[N], H[M]\}&=&H_{a}[ND^aM-MD^aN]+C[\frac{2D_a\phi}{(D-1)\phi}(ND^aM-MD^aN)], \label{HH}\\
\{C[\lambda], H[M]\}&=&H[\frac{\lambda M}{2}]+\ints
N\lambda\sqrt{h}(-\frac{D+1}{2} V(\phi)+\frac{D-1}{2}\phi V'(\phi)). \label{Sc} \ea
The Poisson bracket (\ref{Sc}) implies a new secondary constraint for the consistency of the constraint $C$ during evolution as:
\ba -\frac{D+1}{2} V(\phi)+\frac{D-1}{2}\phi V'(\phi)=0.
\label{equationofV}\ea
Further calculations show that this constraint is of
second-class and hence has to be solved. In the vacuum case we have following two different solutions for Eq. (\ref{equationofV})\footnote{The case of non-dynamical $\phi$ is not included, since we consider the STT different from GR (plus a cosmological constant).}:
\ba
 V(\phi)=0 \quad or \quad  V(\phi)=c\phi^{\frac{D+1}{D-1}}, \label{solcon}
\ea
 where $c$ is some undetermined
dimensional constant. Thus the consistency condition requires that we can only have two special forms of potentials
when $\omega(\phi)=-\frac{D}{D-1}$. With these two solutions, the set $( H,H_a,C)$ also comprise a first-class system and the action
(\ref{action}) become  invariant under the following conformal
transformations:
\ba g_{\mu\nu}\rightarrow e^\lambda
g_{\mu\nu}, \quad \phi\rightarrow e^{-\frac{D-1}{2}\lambda}\phi. \label{conformalt}
\ea
The geometrical meaning of the conformal
constraint (\ref{sc}) can be understood by its actions on the phase
space variables:
\ba &&\{h_{ab}, C(\lambda)\}=\lambda h_{ab}, \quad
\{\Pi^{ab}, C(\lambda)\}=-\lambda \Pi^{ab},  \\
&&\{\phi, C(\lambda)\}=-\frac{D-1}{2}\lambda \phi, \quad \{\pi, C(\lambda)\}=\frac{D-1}{2}\lambda
\pi. \ea
Obviously the above transformations agree
with the spacetime conformal transformations
(\ref{conformalt}). Moreover, due to this additional constraint,
the physical degrees of freedom in this special sector are
equal to those of GR in D+1 dimensions. Finally, since the initial value problem in this special
sector is a very subtle issue \cite{So, olmo},  we leave the comparison
between the Hamiltonian formulation and the Lagrangian formulation
for future study.

\section{CONNECTION-DYNAMICAL FORMULATION}
\subsection{Review of the connection dynamics for GR in $D+1$ dimensions}
In this subsection, we will give a brief introduction to the approach in Ref.\cite{Th03} for constructing the connection dynamics of GR in arbitrary dimensions. The framework will be employed to formulate the connection dynamics of STT in $D+1$ dimensions in the next subsection.

As is well known,  the ADM Hamiltonian formulation
of vacuum $(D+1)$-dimensional GR is based on a phase space coordinatised by a canonical pair $(h_{ab}, P^{ab})$ with Poisson brackets
\ba \{h_{ab}(x), P^{cd}(y)\}&=&\delta^{c}_{(a}\delta^{d}_{b)}\delta^{D}(x, y), \quad\{h_{ab}(x), h_{cd}(y)\}=\{P^{ab}(x), P^{cd}(y)\}=0. \label{admbrac}\ea
The spatial diffeomorphism constraint and Hamiltonian constraint for Lorentzian spacetime read respectively
\ba
V_{a}&=&-2h_{ac}D_{b}P^{bc}, \label{vector}\\ H&=&-\frac{1}{2}\sqrt{det(h)}R^{(D)}+\frac{2}{\sqrt{det(h)}}(h_{ac}h_{bd}-\frac{1}{D-1}h_{ab}h_{cd})P^{ab}P^{cd}. \label{admhamiltonian}
\ea
To formulate GR in terms of a gauge theory, the central idea is to extend the ADM phase space by additional degrees of freedom and then impose additional first-class constraints such that£¬ after symplectic reduction with respect to these constraints, we can recover the original ADM phase space. The canonical pair of the extended phase space consists of a Lie algebra valued one form $A_{aIJ}$ with dimension $N$ and the corresponding conjugate momentum $\pi^{aIJ}$ which is a Lie algebra valued weight-one vector density.

 It is argued in \cite{Th03} that the underlying gauge which one should choose without gauge fixing is $SO(1, D)$ or $SO(D+1)$ for $(D+1)$-dimensional spacetime, and an additional constraint will appear due to the mismatching between the number of the degrees of freedom of $(A_{aIJ}, \pi^{aIJ})$ and that of $(h_{ab}, P^{ab})$ modulo the above constraints. In practical terms, the degrees of freedom for $A_{a}{}^{\IJ}$ are $\frac{D^{2}(D+1)}{2}$ where $\{a\in 1...D\}$ and $\{I, J\in0. . . D\}$. Note that the two internal indices are antisymmetric with each other, and hence contribute $\frac{D(D+1)}{2}$ degrees of freedom. After subtracting the number of Gaussian constraints, $\frac{D(D+1)}{2}$, and the degrees of freedom of $h_{ab}$, $\frac{D(D+1)}{2}$, the remaining degrees of freedom read $\frac{D^{2}(D+1)}{2}-\frac{D(D+1)}{2}-\frac{D(D+1)}{2}=\frac{D(D-2)(D+1)}{2}$,
 which means there are $\frac{D(D-2)(D+1)}{2}$ additional constraints. These constraints could be imposed on the momentum $\pi^{aIJ}$ conjugate to $A_{aIJ}$, if we require $\pi^{aIJ}$ be determined by the co-D-bein $e_{a}^{I}$. Since $\pi^{aIJ}$ has degrees of freedom $\frac{D^{2}(D+1)}{2}$, while $e_{a}^{I}$ has only $D(D+1)$, the subtraction $\frac{D^{2}(D+1)}{2}-D(D+1)=\frac{D(D-2)(D+1)}{2}$ exactly matches with the number of the desired remaining constraints. Thus we expect to build $\pi^{aIJ}\propto n^{[I}E^{aJ]}$ on this new constraint surface, where $E^{aJ}:=\sqrt{det(h)}h^{ab}e^{I}_{b}$,  $h^{ab}$ is the inverse of $h_{ab}\equiv e_{a}^{I}e_{bI}$, $n^{I}$ is the internal vector orthogonal to $e_{a}^{I}$ and uniquely determined (up to a sign) by $e_{a}^{I}$ through
\ba
n_{I}:=\frac{1}{D!}\frac{1}{\sqrt{det(h)}}\epsilon^{a_{1}. . . a_{D}}\epsilon_{IJ_{1}. . . J_{D}}e_{a_{1}}^{J_{1}}. . e_{a_{D}}^{J_{D}}. \label{nI}
\ea
Note that one has $ n_{I}n^{I}=1$ for $SO(D+1)$ and $ n_{I}n^{I}=-1$ for $SO(1, D)$. In the following, we will choose the compact gauge group $SO(D+1)$ and require that
\ba
\pi^{aIJ}:=2\sqrt{det(h)}h^{ab}n^{[I}e^{J]}_{b}=2n^{[I}E^{|a|J]}, \label{piaIJ}
\ea
 on the constraint surface of ``Simplicity Constraint". It should be noted that $\frac{D(D-2)(D+1)}{2}=0$ for $D=2$ and hence no simplicity constraint is needed under this case.

 To get an explicit expression of the simplicity constraint, for any given unit internal vector $n_{I}$, we define $E^{aI}:=-\pi^{aIJ}n_{J}$ and its corresponding quantities:
  \ba
  Q^{ab}:=E^{a}_{I}E^{b}_{J}\eta^{IJ}, Q_{ac}Q^{cb}:=\delta ^{b}_{a}, E_{a}^{J}:=Q_{ab}E^{bI},\label{Eaj}
 \ea
 where $\eta^{IJ}$ is the internal metric.
 Furthermore, we define the transversal projector:
 \ba
 \bar{\eta}^{I}_{J}[n]:=\delta^{I}_{J}-n^{I}n_{J}. \label{etaij}
 \ea
Using $ E_{a}^{J}$ and $\bar{\eta}^I_{J}$, we can define the tracefree and transverse projector:
\ba
P^{aIJ}_{bKL}[E]:=\delta ^{a}_{b}\bar{\eta}^{I}_{[K}\bar{\eta}^{J}_{L]}-\frac{2}{D-1}E^{a[I}E_{b[K}\bar{\eta}^{J]}_{L]}. \label{Paijbkl}
\ea
 Next we define
\ba
\bar{\pi}^{aIJ}_{T}:=P^{aIJ}_{bKL}\pi^{bKL}=\bar{\pi}^{aIJ}+\frac{2}{D-1}\bar{\pi}^{[I}E^{|a|J]}. \label{pibart}
\ea
Note that $\bar{\pi}^{aIJ}_{T}$ satisfies $E_{aI}\bar{\pi}^{aIJ}_{T}=0$ and $\bar{\pi}^{aIJ}_{T}n_{I}=0$.  The key observation is that $\bar{\pi}^{aIJ}_{T}$ has only $\frac{D(D-2)(D+1)}{2}$ degrees of freedom which is just the number of degrees of freedom we need to remove. Hence a given tensor $\pi ^{aIJ}$ can be decomposed into three parts:
 \ba
\pi ^{aIJ}=\bar{\pi}^{aIJ}_{T}-\frac{2}{D-1}\bar{\pi}^{[I}E^{|a|J]}+2n^{[I}E^{|a|J]}, \label{piaijdec}
 \ea
  where $\bar{\pi}^{J}:=E_{aI}\bar{\pi}^{aIJ}$, and hence $\bar{\pi}^{[I}E^{|a|J]}$ is normal to $n^{I}$ but not normal to $E^{aI}$. On the other hand, as shown in Ref.\cite{Th03}, one can always choose a suitable internal vector $n_{I}$ such that
\ba
\bar{\pi}^{J}[\pi , n]=\bar{\pi}^{aIJ}[\pi, n]Q_{ab}[\pi, n]E^{b}_{I}=0. \label{pibarj}
\ea
Thus, by employing the chosen $n^I$,
one obtains an intrinsic decomposition: $\pi ^{aIJ}= \bar{\pi}^{aIJ}_{T}+2n^{[I}E^{|a|J]}$. Hence one would like to impose the simplicity constraint as the necessary and sufficient condition for a vanishing $\bar{\pi}^{aIJ}_T$. Let $D\geq 3$ and
\ba
S^{ab}_{\bar{M}}:=\frac{1}{4}\epsilon_{I_{0}I_{1}I_{2}I_{3}\bar{M}}\pi^{aI_{0}I_{1}}\pi^{bI_{2}I_{3}}, \label {simp}
\ea
where $\bar{M}$ is any totally skew $(D-3)$-tuple of indices in ${0, 1, . . , D}$ , which stands for the set of the other $(D-3)$ antisymmetric indices $\{I_{4}, I_{5}, . . , I_{D}\}$ . Then for any unit vector $n^I$, one has \cite{Th03}
\ba S^{ab}_{\bar{M}}=0 , \forall \bar{M}, a, b \Leftrightarrow P^{aIJ}_{bKL}[\pi , n]\pi ^{bKL}=0. \label {Sab} \ea
Therefore the desired simplicity constraint reads $S^{ab}_{\bar{M}}=0$.

 Now we consider the Hamiltonian formalism of a $SO(D+1)$ gauge theory with connection $A_{aIJ}$ and its conjugate momentum $\pi ^{bKL}$ as basic variables. These variables are subject to the Poisson brackets
\ba \{A_{aIJ}(x), \pi ^{bKL}(y)\}=4\beta \delta ^{a}_{b}\delta ^{K}_{[I}\delta ^{L}_{J]}\delta ^{D}(x,y), \quad\{A_{aIJ}(x), A_{bKL}(y)\}=\{\pi ^{aIJ}(x), \pi ^{bKL}(y)\}=0, \label {Poisson1}\nn\\ \ea
where $\beta$ is the ``Immirzi-like parameter" (it is structurally different from the Immirzi parameter in $D=3$) in $D$ dimensions. Then the Gaussian constraint and simplicity constraint read respectively \cite{Th03}:
\ba
G^{IJ}&:=&\mathcal {D}_{a}\pi ^{aIJ}:=\partial _{a}\pi ^{aIJ}+2A_{a}^{[I}{}_{K}\pi ^{a|K|J]}, \label {Gauss} \\
S^{ab}_{\bar{M}}&:=&\frac{1}{4}\epsilon_{IJKL\bar{M}}\pi^{aIJ}\pi^{bKL}. \label {simplicity} \ea
The ADM variables can be related to the Yang-Mills variables by the following map,
\ba
hh^{ab}&:=&\frac {1}{2}\pi ^{aIJ}\pi^b{}_{IJ} , \label {symred1}\\
P^{ab}&:=&\frac {1}{8\beta}\Big(h^{a[c}[A_{cIJ}-\Gamma _{cIJ}\pi ^{b]IJ}+h^{b[c}[A_{cIJ}-\Gamma _{cIJ}]\pi ^{a]IJ}\Big) \nn\\
&=:&\frac{1}{4}h^{d(a}K_{cIJ}\pi ^{[b)IJ}\delta ^{c]}_{d}, \label {symred2}
\ea
where $\Gamma _{cIJ}(\pi)$ satisfies ($\approx$ means vanishing on the simplicity constraint surface)
\ba D_{a}\pi^{bIJ}:=\partial_{a}\pi^{bIJ}+\Gamma^{b}_{ac}\pi^{cIJ}+2\Gamma_{a}{}^{[I}{}_{K}\pi^{|bK|J]}-\Gamma^{c}_{ac}\pi^{bIJ}\approx0. \label{Gammacij}
\ea
Eq.(\ref{Gammacij}) can be explicitly solved as
\ba
\Gamma_{aIJ}[\pi]:=\frac{2}{D-1}T_{aIJ}+\frac{D-3}{D-1}\bar{T}_{aIJ}+\Gamma^{b}_{ac}T^{c}_{bIJ}, \label{Gamma}
\ea
where $T_{aIJ}:=\pi_{bK[I}\partial_{a}\pi^{bK}{}_{J]}$, $\bar{T}_{aIJ}:=\bar{\eta}^{K}_{I}\bar{\eta}^{L}_{J}T_{aKL}$, $T^{c}_{bIJ}:=\pi_{bK[I}\pi^{cK}{}_{J]}$, and $\Gamma ^{b}_{ac}$ is the Levi-Civita connection compatible with $h_{ab}$. Using $\pi^{aIJ}\approx2n^{[I}E^{|a|J]}$, one can show that $\Gamma_{aIJ}[\pi]$ is compatible with $e^{I}_{a}$, i.e.,
\ba D_{a}e_{b}^{I}=\partial_{a}e^{I}_{b}-\Gamma^{c}_{ab}e^{I}_{c}+\Gamma^{IJ}_{a}e_{bJ}=0, \label{gammae}\ea
on the simplicity constraint surface. It was shown in Ref.\cite{Th03} that using the symplectic structure (\ref{Poisson1}), one can correctly recover
 \ba \{h_{ab}(x), P^{cd}(y)\}=\delta^{c}_{(a}\delta^{d}_{b)}\delta^{D}(x, y), \quad\{h_{ab}(x), h_{cd}(y)\}=\{P^{ab}(x), P^{cd}(y)\}=0,
 \ea
on the simplicity and Gaussian constraints surface. Hence, the map defined by Eqs. (\ref{symred1}) and (\ref{symred2}) gives a symplectic reduction from the Yang-Mills phase space to the ADM phase space.
The diffeomorphism constraint (\ref{vector}) and Hamiltonian constraint (\ref{admhamiltonian}) can be expressed in terms of the new variables as
\ba
\mathcal{V}_{a}&=&\frac {1}{2\beta}F_{abIJ}\pi ^{bIJ},\label {vectornew}\\
\mathcal{H}&=&\frac{1}{2\sqrt{h}}\Big(F_{abIJ}\pi^{aIK}\pi^b{}_{K}{}^J+4\bar{D}^{aIJ}_T(F^{-1})_{aIJ, bKL}\bar{D}^{bKL}_T
+\frac{1}{(D-1)^2}[D_b{}^aD_a{}^b-(D_a{}^a)^2]\Big)\nn\\
&&+\frac{1}{8\beta^2(D-1)^2\sqrt{h}}[D_b{}^aD_a{}^b-(D_a{}^a)^2], \label{scalarnew}
\ea
where $F_{abIJ}\equiv2\partial_{[a}A_{b]IJ}+2A_{a[I|K|}A_{|b|}{}^{K}{}_{J]}$ is the curvature of $A_{aIJ}$. Here we defined
\ba
D_{b}{}^{a}&:=&\pi^{aK}{}_{J}(D_{b}\pi^{cJL})\pi_{cKL}, \label {Dab}\\
 (F^{-1})_{aIJ, bKL}&:=&\frac{1}{4}[Q_{ab}\bar{\eta}_{K[I}\bar{\eta}_{J]L}-2E_{b[I}\bar{\eta}_{J][K}E_{aL]}], \label{Fminus}\\
D^{aIJ}&:=&\pi^{b[I}{}_{K}\mathcal{D}_{b}\pi^{a|K|J]}, \label{Daij}
\ea
where $\mathcal{D}_{a}$ is the covariant differential of $A$ acting only on internal indices, i.e.,
\ba
\mathcal{D}_{a}\pi^{bIJ}:=\partial _{a}\pi ^{bIJ}+2A_{a}^{[I}{}_{K}\pi ^{b|K|J]},
\ea
and $\bar{D}^{aIJ}_T$ is the tracefree and transverse part of $D^{aIJ}$ defined by
\ba
\bar{D}^{aIJ}_T&:&=P^{aIJ}_{bKL}\cdot D^{bKL}. \label{DbaraijT}
\ea
All of the constraints (\ref{Gauss}),(\ref{simplicity}),(\ref{vectornew}),(\ref{scalarnew}) are proved to be of first class \cite{Th03}.

\subsection{Connection dynamics for STT in $D+1$ dimensions}
It was recently shown in Ref.\cite{Zh11c} that the STT in 3+1 dimensions can be cast into connection-dynamical formalism. However, a
connection-dynamical formalism for STT in arbitrary dimensions is still lacking. Thus our task now is to extend the approach introduced in the last subsection to formulate a connection dynamics of GR to $(D+1)$-dimensional STT. Recall that in order to build the connection dynamics of $(D+1)$-dimensional GR, we need to define the suitable canonical variables $\pi^{aIJ}$ and $A_{aIJ}$ of Yang-Mills fields and then construct the ADM phase space by symplectic reduction. For STT, the question becomes how to get the ADM-like phase space obtained in section \Rmnum 2 by a suitable symplectic reduction of a $so(D+1)$ Yang-Mills phase space.  Note that, besides Yang-Mills variables, we also need a scalar field and its momentum. Hence the phase space of the gauge theory consists of the canonical pairs $(\tilde{ A}_{aIJ}, \pi^{aIJ})$ and $(\phi, \pi)$, with basic Poisson brackets
\ba
\{\tilde{A}_{aIJ}(x), \pi ^{bKL}(y)\}&=&4\beta \delta ^{a}_{b}\delta ^{K}_{[I}\delta ^{L}_{J]}\delta ^{D}(x,y), \quad\{\tilde{A}_{aIJ}(x), \tilde{A}_{bKL}(y)\}=\{\pi ^{aIJ}(x), \pi ^{bKL}(y)\}=0, \nn\\
\{\phi(x), \pi(y)\}&=&\delta^D(x, y),    \quad\quad\quad\quad\quad\quad\{\phi(x), \phi(y)\}=\{\pi(x), \pi(y)\}=0.
\label {Poisson2}
\ea
To construct the ADM variables from the Yang-Mills variables, we first define
\ba
\beta\tilde{K}_{aIJ}=\tilde{A}_{aIJ}-\tilde{\Gamma}_{aIJ},
\ea
where $\beta$ is an arbitrary real number, and
\ba
\tilde{\Gamma}_{aIJ}:=\Gamma_{aIJ}[\pi;x)+S_{aIJ}[\pi;x),
\ea
where $S_{aIJ}$ refers to certain function vanishing on the simplicity constraint surface, and $\Gamma_{aIJ}$ is defined by Eq.(\ref{Gamma}). As shown in \cite{Th03}, $\tilde{\Gamma}_{aIJ}$ can be chosen as the functional derivative of a generating function $F[\pi]$ such that $\tilde{K}_{aIJ}$ commutes with itself in Poisson brackets. This property will simplify the calculations of our constraint algebra.
Then we define a map from the phase space of the gauge field coupled with the scalar field to ADM phase space of STT by
\ba
\sqrt{h}h^{ab}&:=&\frac {1}{2}\pi ^{aIJ}\pi^b{}_{IJ} , \label {symred21}\\
\Pi^{ab}&:=&\frac{1}{4}h^{d(a}\tilde{K}_{cIJ}\pi ^{[b)IJ}\delta ^{c]}_{d}, \label {symred22}\\
\phi&:=&\phi,\label{symred23}\\
\pi&:=&\pi.\label{symred24}
\ea
Note that the Gaussian constraint of the gauge theory reads
\ba
G^{IJ}&:=&\mathcal {\tilde{D}}_{a}\pi ^{aIJ}=\partial _{a}\pi ^{aIJ}+2\tilde{A}_{a}^{[I}{}_{K}\pi ^{a|K|J]}, \label {Gauss1}\ea
while the simplicity constraint keeps the same form as Eq.(\ref {simplicity}).
Now it is straightforward to check  that $h_{ab}[\pi]$ and $\Pi^{ab}[A, \pi]$ defined in Eqs. (\ref{symred21}) and (\ref{symred22}) are Dirac observables with respect to the Gaussian and simplicity constraints and obey the standard Poisson brackets :
\ba\{h_{ab}(x), \Pi^{cd}(y)\}=\delta^{c}_{(a}\delta^{d}_{b)}\delta^{D}(x, y), \quad\{h_{ab}(x), h_{cd}(y)\}=\{\Pi^{ab}(x),\Pi^{cd}(y)\}=0.
\label{admbrackets}
 \ea
Therefore the map defined by Eqs. (\ref{symred21}), (\ref{symred22}), (\ref{symred23}) and (\ref{symred24}) is also a symplectic reduction.

To reformulate the geometrical dynamics of $(D+1)$-dimensional STT by connection dynamics. We first consider the sector of $\omega(\phi)\neq-\frac{D}{D-1}$. Eq.(\ref{symred22}) implies that $\tilde{K}_{a}{}^b:=\frac{1}{4}\tilde{K}_{aIJ}\pi ^{bIJ}$ is related to the extrinsic curvature in Eq.(\ref{excur}) by
\ba
\tilde{K}_{a}{}^b
&=&\phi
\sqrt{h}K_{a}{}^b+\frac{2(\frac{D-1}{2}\phi\pi-\Pi)\delta_a^b}{(D-1)(D+(D-1)\omega)}. \label{Ktilde}
\ea
Straightforward calculations show that the original diffeomorphism and Hamiltonian constraints (\ref{dc}) and (\ref{hc}) can be respectively written in terms of new variables as
\ba
\mathcal{H}_a&=&\frac{1}{2\beta}\tilde{F}_{abIJ}\pi^{bIJ}+\pi\partial_a\phi, \label{stvector}\\
\mathcal{H}&=&\frac{\phi}{2\sqrt{h}}\Big(\tilde{F}_{abIJ}\pi^{aIK}\pi^b{}_{K}{}^J+4\tilde{\bar{D}}^{aIJ}_T(\tilde{F}^{-1})_{aIJ, bKL}\tilde{\bar{D}}^{bKL}_T
+\frac{1}{(D-1)^2}[\tilde{D}_b{}^a\tilde{D}_a{}^b-(\tilde{D}_a{}^a)^2] \Big) \nn\\
&&+\frac{1}{8\phi\beta^2(D-1)^2\sqrt{h}}[\tilde{D}_b{}^a\tilde{D}_a{}^b-(\tilde{D}_a{}^a)^2]+\frac{(\tilde{D}_a{}^a
-2\beta(D-1)\phi\pi)^2}{8\phi\beta^2(D-1)(D+(D-1)\omega)\sqrt{h}}\nn\\
&&+\frac12\sqrt{h}\Big(\frac{\omega(\phi)}{\phi}(\tilde{D}_a\phi)
\tilde{D}^a\phi+2\tilde{D}_a\tilde{D}^a\phi+2 V(\phi)\Big), \label{stscalar}
\ea
where $\tilde{F}_{abIJ}\equiv2\partial_{[a}\tilde{A}_{b]IJ}+2\tilde{A}_{a[I|K|}\tilde{A}_{|b|}{}^{K}{}_{J]}$ and the definitions for $\tilde{D}_a{}^b$, $(\tilde{F}^{-1})_{aIJ, bKL}$ and $\tilde{\bar{D}}^{aIJ}_T$ take the same forms as Eqs. (\ref{Dab}), (\ref{Fminus}) and (\ref{DbaraijT}) except for $A_{aIJ}$ being replaced by $\tilde{A}_{aIJ}$, and the generalized derivative satisfies $\tilde{D}_{a}\pi^{bIJ}:=\partial_{a}\pi^{bIJ}+\Gamma^{b}_{ac}\pi^{cIJ}+2\tilde{\Gamma}_{a}{}^{[I}{}_{K}\pi^{|bK|J]}-\Gamma^{c}_{ac}\pi^{bIJ}= 0$ on the simplicity constraint surface. The total Hamiltonian now become
\ba
H_{tot}=\int_{\Sigma}d^Dx\Big(c^{\bar{M}}_{ab}S^{ab}_{\bar{M}}+\frac{1}{2}f_{IJ}G^{IJ}+N^{a}\mathcal{H}_{a}+N\mathcal{H}\Big). \label{totalham}\ea
It is easy to check that the smeared Gaussian constraint $G[f]:=\int_{\Sigma}d^{D}x\frac{1}{2}f_{IJ}(x)G^{IJ}(x)$ generates $SO(D+1)$ gauge transformations on the phase space as
\ba
\Big\{\tilde{A}_{aIJ}, G[f]\Big\}=-2\beta \mathcal{\tilde{D}}_{a}f_{IJ}, \quad
\Big\{\pi^{aIJ}, G[f]\Big\}=2\beta[f, \pi^a]^{IJ}.
\ea
The smeared diffeomorphism constraint $\mathcal{\tilde{H}}[\overrightarrow{N}]:=\int_{\Sigma}d^DxN^a(\mathcal{H}_{a}-\frac{1}{2\beta}\tilde{A}_{aIJ}G^{IJ})$ generates the spatial diffeomorphism transformations on the phase space as
\ba
\Big\{\tilde{A}_{aIJ}, \mathcal{\tilde{H}}[\overrightarrow{N}]\Big\}&=&2\mathcal{L}_{\vec{N}}\tilde{A}_{aIJ}, \quad
\Big\{\pi^{aIJ}, \mathcal{\tilde{H}}[\overrightarrow{N}]\Big\}=2\mathcal{L}_{\vec{N}}\pi^{aIJ},\\
\Big\{\phi,\mathcal{\tilde{H}}[\overrightarrow{N}]\Big\}&=&\mathcal{L}_{\vec{N}}\phi,\quad\quad\quad\Big\{\pi,\mathcal{\tilde{H}}
[\overrightarrow{N}]\Big\}=\mathcal{L}_{\vec{N}}\pi.
\ea
Thus we can show that the constraint algebra has the following Poisson subalgebra:
\ba
\Big\{G[f], S^{ab}_{\bar{M}}[d^{\bar{M}}_{ab}]\Big\}&=&S^{ab}_{\bar{M}}[\sum^{D-3}_{i=1}2\beta f^{I_{i}}{}_{I'_{i}}d^{I_{1}. . . I'_{i}. . . I_{D-3}}_{ab}], \label{simpgauss}\\
\Big\{S^{ab}_{\bar{M}}[c^{\bar{M}}_{ab}], S^{ab}_{\bar{M}}[d^{\bar{M}}_{ab}]\Big\}&=&0, \label{simpsimp}\\
\Big\{G[f], G[g]]\Big\}&=&-2\beta G[f, g], \label{GaussGauss}\\
\Big\{G[f], \mathcal{\tilde{H}}[\overrightarrow{N}]\Big\}&=&G[-\mathcal{L}_{\vec{N}}f], \\
\Big\{S^{ab}_{\bar{M}}[c^{\bar{M}}_{ab}], \mathcal{\tilde{H}}[\overrightarrow{N}]\Big\}&=&2S^{ab}_{\bar{M}}[\mathcal{L}_{\overrightarrow{N}}c^{\bar{M}}_{ab}], \label{simph}\\
\Big\{\mathcal{\tilde{H}}[\overrightarrow{M}], \mathcal{\tilde{H}}[\overrightarrow{N}]\Big\}&=&2\mathcal{\tilde{H}}([\overrightarrow{M}, \overrightarrow{N}]). \label{hh}
\ea
To simplify the calculation of the Poisson brackets, we notice that the simplicity constraint commutes with itself as well as Gaussian and diffeomorphism constraints. Thus we can rewrite the Hamiltonian constraint modulo the simplicity constraint as
\ba
\tilde{\mathcal{H}}
&=&\frac{1}{2\sqrt{h}\phi}(\tilde{K'}_{aIJ}\pi^{bIJ}\tilde{K'}_{bKL}\pi^{aKL}-\tilde{K'}_{aIJ}\pi^{aIJ}\tilde{K'}_{bKL}\pi^{bKL})
-\frac12\phi\sqrt{h}R^{(D)}\nn\\
&&+\frac{1}{2(\frac{D}{D-1}+\omega)\phi\sqrt{h}}(\tilde{K'}_{aIJ}\pi^{aIJ}+\pi\phi)^2+\frac{\omega}{2\phi}\sqrt{h}(D_a\phi)
D^a\phi+\sqrt{h}D_aD^a\phi+\sqrt{h} V(\phi),\nn\\
\label{newhamil}
\ea
where $\tilde{K}'_{aIJ}:=\frac{1}{4}\tilde{K}_{aIJ}$. Now we can show that the Poisson brackets between the smeared Hamiltonian constraint with itself and other constraints are also closed as
\ba
\Big\{G[f], \mathcal{\tilde{H}}[M]\Big\}&=&0, \label{GaussH}\\
\Big\{\mathcal{\tilde{H}}[\overrightarrow{N}], \mathcal{\tilde{H}}[M]\Big\}&=& \mathcal{\tilde{H}}[L_{\overrightarrow{N}}M], \label{hH}\\
\Big\{S^{ab}_{\bar{M}}[c^{\bar{M}}_{ab}], \mathcal{\tilde{H}}[M]\Big\}&=&0, \label{simpH}\\
\Big\{\mathcal{\tilde{H}}[M], \mathcal{\tilde{H}}[N]\Big\}&=& \int_\Sigma d^Dx\frac{1}{2h}\pi^{aIK}\pi^{b}{}_{K}{}^{J}(ND_bM-MD_bN)\mathcal{H}_a \nn\\&&\qquad+\frac{1}{4\beta h} [\pi^aD_aN, \pi^bD_bM]^{IJ}{G}_{IJ}. \label{HH}
\ea
The detailed calculation of Eq.(\ref{HH}) will
be presented in Appendix A. Hence all the constraints now are of first class. To summarize, the STT of gravity in the
sector $\omega(\phi)\neq \frac{D}{D-1}$ have been cast into the
$so(D+1)$-connection dynamical formalism with a first-class constraint system.

Next we consider the other sector of $\omega(\phi)= -\frac{D}{D-1}$. Besides the diffeomorphism and Hamiltonian constraints, the geometrical dynamics of STT contains an extra primary conformal constraint (\ref{sc}). On the phase space of the gauge field coupled with the scalar field, the total Hamiltonian can be expressed as a liner combination
\ba
H_{tot}=\int_{\Sigma}d^Dx\Big(c^{\bar{M}}_{ab}S^{ab}_{\bar{M}}+\lambda C+\frac{1}{2}f_{IJ}G^{IJ}+N^{a}\mathcal{H}_{a}+N\mathcal{H}\Big), \label{htotal1}
\ea
 where the simplicity, Gaussian and diffeomorphism constraints keep the same form as Eqs. (\ref{simplicity}), (\ref{Gauss1}) and (\ref{stvector}), while the conformal and Hamiltonian constraints read respectively:
\ba
C&=&\frac{1-D}{2}(\tilde{K}_a{}^a+\pi\phi),\label{Snew}\\
\mathcal{H}&=&\frac{\phi}{2\sqrt{h}}\Big(\tilde{F}_{abIJ}\pi^{aIK}\pi^b{}_{K}{}^J+4\tilde{\bar{D}}^{aIJ}_T(\tilde{F}^{-1})_{aIJ, bKL}\tilde{\bar{D}}^{bKL}_T+\frac{1}{(D-1)^2}[\tilde{D}_b{}^a\tilde{D}_a{}^b-(\tilde{D}_a{}^a)^2]\Big)\nn\\
&&+\frac{1}{8\phi\beta^2(D-1)^2\sqrt{h}}[\tilde{D}_b{}^a\tilde{D}_a{}^b-(\tilde{D}_a{}^a)^2]
+\frac12\sqrt{h}\Big(\frac{-D}{(D-1)\phi}(\tilde{D}_a\phi)\tilde{D}^a\phi+2\tilde{D}_a\tilde{D}^a\phi+2 V(\phi)\Big).\nn\\ \label{ham1}
\ea
 After solving the second-class constraint as shown in section \Rmnum 2, straightforward calculations show that the constraint algebra is still closed as:
\ba
\{S^{ab}_{\bar{M}}[d^{\bar{M}}_{ab}], C[\lambda]\}&=&S^{ab}_{\bar{M}}[(D-1)d^{\bar{M}}_{ab}],\label{SabC} \\
\{G[f], C[\lambda]\}&=&0, \\
\{C[\lambda], \mathcal{\tilde{H}}[M]\}&=&\mathcal{\tilde{H}}[\frac{\lambda M}{2}], \\
\{\mathcal{\tilde{H}}[N], \mathcal{\tilde{H}}[M]\}
&=&\int_\Sigma d^Dx\frac{1}{2h}\pi^{aIK}\pi^{b}{}_{K}{}^{J}(ND_bM-MD_bN)\mathcal{H}_a+\frac{[\pi^aD_aN, \pi^bD_bM]^{IJ}}{4\beta h}{G}_{IJ}\nn\\
&&\quad+C[\frac{2D_a\phi}{(D-1)\phi}(ND^aM-MD^aN)]. \label{HH1}\ea
where we have rewritten the smeared Hamiltonian constraint corresponding to (\ref{ham1}) into the following equivalent form modulo the simplicity constraint:
\ba
\tilde{\mathcal{H}}[M]
&=&\int_\Sigma d^DxM\Big[\frac{1}{2\sqrt{h}\phi}(\tilde{K'}_{aIJ}\pi^{bIJ}\tilde{K'}_{bKL}\pi^{aKL}-\tilde{K'}_{aIJ}\pi^{aIJ}\tilde{K'}_{bKL}\pi^{bKL})
-\frac12\phi\sqrt{h}R^{(D)}\nn\\
&&\qquad-\frac{D}{2(D-1)\phi}\sqrt{h}(D_a\phi)
D^a\phi+\sqrt{h}D_aD^a\phi+\sqrt{h} V(\phi)\Big].\nn \\\ea
The derivation of Eq.(\ref{SabC}) will be given in Appendix A. Obviously the Poisson brackets among the other constraints are also weakly equal to zero. Hence all the constraints in this case are also of first class. To summarize, the STT of gravity in both sectors of $\omega(\phi)\neq-\frac{D}{D-1}$ and $\omega(\phi)=-\frac{D}{D-1}$ have been cast into the $so(D+1)$-connection dynamical formalism with a first-class constraint system.

\section{Concluding Remarks}
As candidate modified gravity theories, STT have received increased attention in issues of ``dark Universe" and nontrivial tests on gravity beyond GR. On the other hand, modern theoretical research explores the possibility of higher dimensional spacetime. In order to study the non-perturbative quantization of higher dimensional STT in LQG scheme, it is necessary to build the connection dynamics of STT in higher spacetime dimensions. The achievements in this
paper are the derivation of the detailed Hamiltonian structure of STT and the construction of their connection dynamics in $(D+1)$-dimensional spacetime. First, by performing Hamiltonian analysis, we derive the Hamiltonian formulation of STT from the $(D+1)$-dimensional Lagrangian formulation in ADM-like variables. Two sectors are marked off by the coupling parameter $\omega(\phi)$. In the sector of $\omega(\phi)\neq -\frac{D}{D-1}$, the canonical structure and constraint algebra of STT are similar to those of GR coupled with a scalar field. In the other sector of $\omega(\phi)= -\frac{D}{D-1}$, the feasible theories are restricted and a new primary constraint generating conformal transformations of
spacetime is obtained. The canonical structure and constraint algebra are also obtained. All the Hamiltonian structures are direct generalization of 4-dimensional case. Next we successfully construct a $so(D+1)$ Hamiltonian connection
formulation of STT in $D+1$ spacetime dimensions, from which the ADM-like Hamiltonian formulation can be obtained by a symplectic reduction. As in higher dimensional GR, a simplicity constraint has to be introduced into the higher dimensional connection dynamics of STT for the symplectic reduction. Finally, we show that the constraint algebra in both sectors of STT are also closed in the connection-dynamical formalism.

It should be noted that we have casted $(D+1)$-dimensional STT into the connection-dynamical formalism with the compact $SO(D+1)$ structure group. Hence it is straightforward to employ the techniques of LQG and those developed in Refs. \cite{Zh11c,At11} to quantize the higher dimensional STT non-perturbatively. This opens the possibility to confront the effects of non-perturbative LQG with those of other higher dimensional quantum gravity theories such as string/M theory.

\begin{acknowledgements}
This work is supported by NSFC (Grant No.11235003 and No. 11305063) and the
Fundamental Research Funds for the Central Universities.
\end{acknowledgements}

\section*{Appendix A}
In this appendix, we will present the detailed calculations for the Poisson brackets (\ref{HH}) and (\ref{SabC}).
First we calculate the Poisson bracket
between two smeared Hamiltonian constraints (\ref{newhamil}). The non-vanishing contributions come only from terms containing derivatives. Hence we
first use $\{\phi(x), \pi(y)\}=\delta^D(x, y)$ to calculate
\ba
&\{&\ints
N\sqrt{h}D_aD^a\phi, \ints\frac{M}{2(\frac{D}{D-1}+\omega)\phi\sqrt{h}}(\tilde{K'}_{aIJ}\pi^{aIJ}+\pi\phi)^2\}_{(\phi, \pi)}-M
\leftrightarrow N \nn\\
&=&\frac{1}{\frac{D}{D-1}+\omega}\ints(ND^aM-MD^aN)D_a(\pi\phi+\tilde{K'}_{bIJ}\pi^{bIJ}), \label{hstart}\ea
and
\ba &\{&\ints \frac{N\sqrt{h}\omega}{2\phi}(D_a\phi)
D^a\phi, \ints\frac{M}{2(\frac{D}{D-1}+\omega)\phi\sqrt{h}}(\tilde{K'}_{bIJ}\pi^{bIJ}+\pi\phi)^2\}_{(\phi, \pi)}-M
\leftrightarrow N \nn\\
&=&\frac{\omega}{\frac{D}{D-1}+\omega}\ints(ND^aM-MD^aN)(\pi\phi+\tilde{K'}_{bIJ}\pi^{bIJ})\frac{D_a\phi}{\phi}. \ea
Note that
\ba
N\sqrt{h}D_aD^a\phi=N\sqrt{h}h^{ab}(\partial_a\partial_b\phi-\Gamma^c_{ab}\partial_c\phi).
\ea
and
\ba
N\sqrt{h}h^{ab}\Gamma^c_{ab}\partial_c\phi&=&\frac
N2\sqrt{h}h^{ab}(\partial_c\phi)\Big(h^{cd}(-\partial_ah_{bd}-\partial_bh_{ad}+\partial_dh_{ab})\Big)\nn\\
&=&\frac
N2\sqrt{h}(\partial_c\phi)\Big(2\partial_a(\frac{\pi^{aIJ}\pi^{c}{}_{IJ}}{2h})-h_{ab}\partial^c(\frac{\pi^{aIJ}\pi^{b}{}_{IJ}}{2h})\Big).
\ea
Therefore, we use
$\Big\{\tilde{K'}_{aIJ}(x), \pi^{bKL}(y)\Big\}=\delta^b_a\eta^{[K}_{I}\eta^{L]}_{J}\delta^D(x,y)$ to calculate
\ba &\Big\{&\ints
N\sqrt{h}(\partial_c\phi)\partial_a(\frac{\pi^{aIJ}\pi^{c}{}_{IJ}}{2h}), \ints\frac{M}{2\sqrt{h}}
\Big(\frac1\phi(\tilde{K'}_{dMN}\pi^{bMN}\tilde{K'}_{bKL}\pi^{dKL}+\frac{2}{(\frac{D}{D-1}+\omega)}\tilde{K'}_{dMN}\pi^{dMN}\pi\nn\\&&-
\frac{\frac{1}{D-1}+\omega}{\frac{D}{D-1}+\omega}\tilde{K'}_{dMN}\pi^{dMN}\tilde{K'}_{bKL}\pi^{bKL})\Big)\Big\}_{(\tilde{K'}, \pi)}-M
\leftrightarrow N \nn\\
&=&\ints\frac14M(\partial_aN)(D_c\phi)\frac{2\pi^{c}{}_{IJ}}{h}\Big(\frac2\phi(\pi^{bIJ}\tilde{K'}_{bKL}\pi^{aKL}-\frac{\frac{1}{D-1}+\omega}
{\frac{D}{D-1}+\omega}\pi^{aIJ}\tilde{K'}_{bKL}\pi^{bKL})
+\frac{2}{(\frac{D}{D-1}+\omega)}\pi^{aIJ}\pi\Big)\nn\\
&&+\frac14M(\partial_aN)(D_c\phi)\frac{\pi^{aIJ}\pi^{c}{}_{IJ}}{h}(\frac{-1}{D-1}\pi_{dKL})\Big(\frac2\phi(\pi^{bKL}\tilde{K'}_{bMN}\pi^{dMN}
-\frac{\frac{1}{D-1}+\omega}{\frac{D}{D-1}+\omega}\pi^{dKL}\tilde{K'}_{bMN}\pi^{bMN})
\nn\\&&+\frac{2}{(\frac{D}{D-1}+\omega)}\pi^{dKL}\pi\Big)-M
\leftrightarrow N,
\ea
and
\ba &\Big\{& \ints-\frac N4\sqrt{h}(\partial_c\phi)
h_{ae}\partial^c(\frac{\pi^{aIJ}\pi^{eIJ}}{h}), \ints\frac{M}{2\sqrt{h}}
\Big(\frac1\phi(\tilde{K'}_{dMN}\pi^{bMN}\tilde{K'}_{bKL}\pi^{dKL}+\frac{2}{(\frac{D}{D-1}+\omega)}\tilde{K'}_{dMN}\pi^{dMN}\pi\nn\\&&-
\frac{\frac{1}{D-1}+\omega}{\frac{D}{D-1}+\omega}\tilde{K'}_{dMN}\pi^{dMN}\tilde{K'}_{bKL}\pi^{bKL})\Big)\Big\}_{(\tilde{K'}, \pi)}-M
\leftrightarrow N \nn\\
&=&\ints-\frac18M\partial^cND_c\phi
h_{ae}\frac{2\pi^{eIJ}}{h}\Big(\frac2\phi(\pi^{bIJ}\tilde{K'}_{bKL}\pi^{aKL}-\frac{\frac{1}{D-1}+\omega}{\frac{D}{D-1}+
\omega}\pi^{aIJ}\tilde{K'}_{bKL}\pi^{bKL})
+\frac{2}{(\frac{D}{D-1}+\omega)}\pi^{aIJ}\pi\Big)\nn\\
&&-\frac18M\partial_aND_c\phi\frac{\pi^{aIJ}\pi^{cIJ}}{2h}(\frac{-2D}{D-1}\pi_{dKL})\Big(\frac2\phi(\pi^{bKL}\tilde{K'}_{bMN}\pi^{dMN}
-\frac{\frac{1}{D-1}+\omega}{\frac{D}{D-1}+\omega}\pi^{dKL}\tilde{K'}_{bMN}\pi^{bMN})
\nn\\&&+\frac{2}{(\frac{D}{D-1}+\omega)}\pi^{dKL}\pi\Big)-
M \leftrightarrow N.\ea
Combination of the above brackets gives
\ba
\ints(ND^aM-MD^aN)\Big(-\frac{\frac{1}{D-1}}{\frac{D}{D-1}+\omega}\pi
D_a\phi-\frac2\phi(\tilde{K'}_{bKL}\pi^{cKL}h_{ac}D^b\phi-\frac{\frac{2}{D-1}+\omega}{2(\frac{D}{D-1}+\omega)}
\tilde{K'}_{bKL}\pi^{bKL}D_a\phi)\Big).\nn\\\ea
Terms containing a derivative in the variation of
$\ints-\frac12\phi N\sqrt{h}R^{(D)}$ read
\ba
&&\ints\frac12\sqrt{h}(-D^aD^b(\phi N)+h^{ab}D_cD^c(\phi N))\delta
h_{ab}\nn\\
&=&\ints\frac12\sqrt{h}(D_aD_b(\phi N)-h_{ab}D_cD^c(\phi N))\delta
(\frac{\pi^{aIJ}\pi^{b}{}_{IJ}}{2h}). \ea
Thus we have
\ba &\Big\{&\ints-\frac12\phi
N\sqrt{h}R^{(D)}, \ints\frac{M}{2\sqrt{h}}\Big(\frac1\phi(\tilde{K'}_{dMN}\pi^{eMN}\tilde{K'}_{eKL}\pi^{dKL}-\frac{\frac{1}{D-1}+\omega}{\frac{D}{D-1}+\omega}
\tilde{K'}_{dKL}\pi^{dKL}\kt_{eMN}\pi^{eMN})
\nn\\&&\quad+\frac{2}{(\frac{D}{D-1}+\omega)}\tilde{K'}_{dMN}\pi^{dMN}\pi\Big)\Big\}
-M \leftrightarrow N \nn\\
&=&\ints(ND_cD^cM-MD_cD^cN)\Big(\frac{1}{(\frac{D}{D-1}+\omega)}\pi-\frac{\frac{1}{D-1}+\omega}{\frac{D}{D-1}+\omega}\tilde{K'}_{aKL}\pi^{aKL}\Big)
\nn\\&&\quad+(ND_cM-MD_cN)D^c\phi\Big(\frac{2}{(\frac{D}{D-1}+\omega)}\pi-\frac{\frac{2}{D-1}+2\omega}{\phi(\frac{D}{D-1}+\omega)}
\tilde{K'}_{aKL}\pi^{aKL}\Big)
\nn\\&&\quad+(ND_aD^bM-MD_aD^bN)\tilde{K'}_{bKL}\pi^{aKL}+(ND_aM-MD_aN)\frac{2D^b\phi}{\phi}\tilde{K'}_{bKL}\pi^{aKL}. \label{hend1}\ea
Combining Eqs. (\ref{hstart})-(\ref{hend1}), we obtain
\ba&&\{\tilde{\mathcal{H}}(N), \tilde{\mathcal{H}}(M)\}\nn\\
&=&\ints(ND_cD^cM-MD_cD^cN)(-\tilde{K'}_{aKL}\pi^{aKL})+(ND^aM-MD^aN)(\pi
D_a\phi)\nn\\&&\quad+(ND_aD^bM-MD_aD^bN)\tilde{K'}_{bKL}\pi^{aKL} \nn\\
&=&\ints(ND^aM-MD^aN)H_a+(D_{a}MD_{b}N-D_{b}MD_{a}N)\tilde{K}^{[ab]}\nn\\
&=&\ints\frac{1}{2h}\pi^{aIK}\pi^{b}{}_{K}{}^{J}(ND_bM-MD_bN)H_a+((D_{a}M)D_{b}N)2\tilde{K}^{[ab]}.\label{tildeHH} \ea
Note that $\tilde{K}^{[ab]}$ is constrained to vanish by the Gaussian and simplicity constraint. To see this, we consider
\ba
G_{IJ}:&=&\mathcal {D}_{a}\pi ^{a}{}_{IJ}=\partial _{a}\pi ^{a}{}_{IJ}+2\tilde{A}_{aK[I}\pi ^{a}{}_{J]}{}^{K}\nn\\
&\approx&-2\beta {\tilde{K}}_{a[I}E^a_{J]}+2\beta \tilde{K}_{[I}n_{J]}=:\bar{G}_{IJ}+2n_{[I}G_{J]},
\ea
where $\tilde{K}_{aI}:=-\tilde{K}_{aLI}n^L$ and $\tilde{K}_I:=K_{aLI}E^{aL}$. It follows that $\tilde{\bar{K}}_I=0$ and $\tilde{K}_{a[I}E^a_{J]}=0$ on the Gaussian constraint surface. Hence we have
\ba
\tilde{K}^{[ab]}E_{aI}E_{bJ}\approx \frac12h^{c[a}\tilde{K}_{cL}E^{b]L}E_{aI}E_{bJ}=\frac{1}{2h}\tilde{K}_{a[I}E^a_{J]}.
\ea
Therefore we have $\tilde{K}^{[ab]}=\frac{1}{4\beta h}\bar{G}_{IJ}E^{aI}E^{bJ}\approx -\frac{1}{4\beta h}G_{IJ}\pi^{aIK}\pi^{b}{}_{K}{}^{J}$ on the simplicity constraint surface. Hence the Poisson bracket (\ref{HH}) can be obtained by Eq.(\ref{tildeHH}).

Next we calculate the Poisson bracket (\ref{SabC}). We notice that the non-vanishing contribution in the conformal constraint coming only from the first term $\tilde{K}_a^a$. Hence we have
\ba
\{S^{bc}_{\bar{M}}[d^{\bar{M}}_{bc}], \frac{1-D}{2}\tilde{K}_a{}^a(y) \}&=&\frac{1-D}{2}\int_\Sigma d^Dx d^{\bar{M}}_{bc}(x)\{S^{bc}_{\bar{M}}(x),\tilde{K}_a^a(y)\}\nn\\
&=&\frac{1-D}{16\beta}\ints d^Dxd^{\bar{M}}_{bc}(x)\epsilon_{ABCD\bar{M}}\pi^{aIJ}\pi^{bAB}\{\pi^{cCD},\tilde{A}_{aIJ}\}\nn\\
&=&S^{ab}_{\bar{M}}[(D-1)d^{\bar{M}}_{ab}].
\ea

\end{document}